\documentclass[12pt]{article}

\usepackage{sbc-template}
\usepackage{graphicx,url}
\usepackage[utf8]{inputenc}
\usepackage{caption}
\usepackage{subcaption}
\usepackage{hyperref}  
\usepackage{xcolor}
\usepackage{changes}
\title{Optimizing Photoplethysmography-Based Sleep Staging Models by Leveraging Temporal Context for Wearable Devices Applications}

\author{
  Joseph A. P. Quino\inst{1,2},
  Diego A. C. Cardenas\inst{1}, \\
  Marcelo A. F. Toledo\inst{1},
  Felipe M. Dias\inst{1,2},\\
  Estela Ribeiro\inst{1},
  José E. Krieger\inst{1} and
  Marco A. Gutierrez\inst{1,2}
}

\address{
  Heart Institute, Clinics Hospital, University of Sao Paulo Medical School\\
  Sao Paulo -- SP -- Brazil
  \nextinstitute
  Polytechnique School, University of Sao Paulo\\
  Sao Paulo --SP -- Brazil
  \email{joseph.pena@hc.fm.usp.br, diego.cardona@hc.fm.usp.br}
  \email{marcelo.arruda@hc.fm.usp.br, f.dias@hc.fm.usp.br}
  \email{estela.ribeiro@hc.fm.usp.br, j.krieger@hc.fm.usp.br}
  \email{marco.gutierrez@incor.usp.br}
}

\begin{document}

\maketitle

\begin{abstract}
Accurate sleep stage classification is crucial for diagnosing sleep disorders and evaluating sleep quality.
While polysomnography (PSG) remains the gold standard, photoplethysmography (PPG) is more practical due to its affordability and widespread use in wearable devices.
However, state-of-the-art sleep staging methods often require prolonged continuous signal acquisition, making them impractical for wearable devices due to high energy consumption.
Shorter signal acquisitions are more feasible but less accurate.
Our work proposes an adapted sleep staging model based on top-performing state-of-the-art methods and evaluates its performance with different PPG segment sizes.
We concatenate 30-second PPG segments over 15-minute intervals to leverage longer segment contexts.
This approach achieved an accuracy of 0.75, a Cohen’s Kappa of 0.60, an F1-Weighted score of 0.74, and an F1-Macro score of 0.60.
Although reducing segment size decreased sensitivity for deep and REM stages, our strategy outperformed single 30-second window methods, particularly for these stages.
\end{abstract}

\begin{resumo}
A classificação precisa do estágio do sono e crucial para diagnosticar distúrbios do sono e avaliar a qualidade do sono.
Embora a polissonografia (PSG) permaneça como padrão ouro, a fotopletismografia (PPG) é mais prática devido à sua acessibilidade e uso generalizado em dispositivos vestíveis.
No entanto, os métodos avançados de classificação do estágio do sono geralmente requerem longos períodos de aquisição contínua prolongada de sinais, tornando-os impraticáveis para dispositivos vestíveis devido ao alto consumo de energia.
Aquisições de sinais mais curtas são mais viáveis, mas menos precisos.
Nosso trabalho propõe um modelo adaptado de classificação do estágio do sono baseado nos métodos avançados de melhor desempenho e avalia o desempenho com diferentes tamanhos de segmentos PPG.
Propomos concatenar segmentos PPG de 30 segundos ao longo de intervalos de 15 minutos para aproveitar contextos de segmentos mais longos.
Essa abordagem atingiu uma precisão de 0,75, um Kappa de Cohen de 0,60, uma pontuação ponderada F1 de 0,74 e uma pontuação F1-Macro de 0,60.
Embora a redução do tamanho dos segmentos tenha diminuído a sensibilidade para os estágios de sono profundo e REM, nossa estratégia superou os métodos de janela única de 30 segundos, especialmente para esses estágios.
\end{resumo}

\section{Introduction}\label{sec:intro}

Considering the potential negative effects of sleep disorders on individual's life, sleep monitoring has become crucial for maintaining health and well-being.
Poor sleep is linked to major health issues such as obesity, diabetes, and cardiovascular disorders \cite{colten_extent_2006}.
Polysomnography (PSG), the gold standard for sleep monitoring, records multiple physiological signals simultaneously during sleep, enabling the identification and analysis of the root causes of sleep disorders \cite{rundo_chapter_2019}.
A key outcome of PSG interpretation is sleep staging or sleep scoring, which assigns labels to each 30-second interval throughout the PSG recordings.
However, PSG-based sleep staging methods are challenging to maintain for continuous monitoring over multiple days due to their high cost and the need for specialized equipment \cite{escourrou_needs_2000}.
As a result, alternative methods for continuous sleep monitoring are gaining popularity.
Photoplethysmography (PPG), widely available in relatively affordable wearable devices, presents a promising alternative for continuous sleep monitoring \cite{djanian_sleep_2022}.

Recent studies have extensively focused on developing strategies to enhance the performance of sleep staging using PPG signals.
As stated in \cite{olsen_flexible_2023}, methods can be roughly categorized based on the PPG segment size required for training and classification.
The state-of-the-art (SOTA) methods validated in publicly available databases, such as the MESA (Multi-Ethnic Study of Atherosclerosis) dataset \cite{zhang_national_2018,chen_racialethnic_2015}, require larger segment sizes.
Models like SleepPPG-Net \cite{kotzen_sleepppg-net_2023} and InsightSleepNet \cite{nam_insightsleepnet_2024} achieve accuracy performance of 0.83 and 0.86, respectively, by leveraging temporal context information over 10 uninterrupted hours of raw PPG signal.

However, in a real application using wearables like smartwatches, such a large data requirement is impractical due to the continuous PPG signal measurement needed, which quickly drains battery life.
A more feasible approach for wearables would be to measure the PPG signal for shorter intervals and periodically repeat the measurement.
Thus, sleep staging methods that work with shorter segment sizes are more suitable for wearable devices.
However, these methods do not perform as well as SOTA methods \cite{song_ai-driven_2023}, with an accuracy of 0.71.

In this study, we propose an adapted version of the SleepPPG-Net \cite{kotzen_sleepppg-net_2023} as our base classification model for sleep staging of four classes: wake, light, deep, and Rapid Eye Movement (REM).
We conducted a series of experiments to evaluate the performance of the adapted SleepPPG-Net using signal segments shorter than 10 hours, aiming to determine the minimum segment length necessary to effectively leverage temporal context information.
To emulate the periodic measurements of wearables and further utilize temporal context, we applied a concatenation strategy before inputting the signal segments into the adapted SleepPPG-Net model.
The proposed approach, adapted to the SleepPPG-Net, outperformed SOTA models that are also able to run with shorter segments of PPG signal.

\section{Materials and Methods} \label{sec:method}
Our methodology consists of the following five steps: (I) Dataset description, which contains the PPG signals and their corresponding labels used for training and validation of the neural network model; (II) Preprocessing, explanation of the processing functions applied to the PPG signal to prepare them to input them into the model; (III) Super Windows description, arranging the PPG signals based on different configurations; (IV) Adapted SleepPPG-Net model, the classification model used for subsequent experiments; and (V) Training and Evaluation, describing the benchmark configurations.

\subsection{Dataset} \label{sub_sec:dataset}
Our experiments were conducted using a sub-set of the Multi-Ethnic Study of Atherosclerosis (MESA) dataset \cite{zhang_national_2018,chen_racialethnic_2015} referred to here as the MESA Sleep dataset.
This dataset contains PSG records of 2,056 subjects of 69.37~$\pm$9.11 years old, with a gender rate of 1:1.2 (males:females).
The MESA Sleep dataset includes PPG signals with a duration of 10.57~$\pm$1.59 hours, recorded using the Nonin 8000J Flex oximeter finger sensor.
Sleep staging followed the ``The AASM Manual for the Scoring of Sleep and Associated Events'' \cite{iber2007aasm}, which assigns a sleep stage to each 30-second interval, referenced as a window.
According to the AASM manual, the possible sleep stages are Wakefulness, NREM1, NREM2, NREM3, and REM.

\subsection{Preprocessing} \label{sub_sec:preprocessing}
\begin{figure}[!htb]
  \centering
  \includegraphics[width=0.7\textwidth]{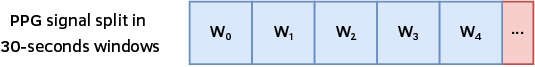}
  \caption{PPG signal split into 30-second windows.}
  \label{fig:ppg_split}
\end{figure}

We used the same preprocessing steps proposed by \cite{almeida_machine-learning_2024,kotzen_sleepppg-net_2023}.
In this study, the raw PPG signal was preprocessed as follows: (I) negation/inversion of the signal; (I) Application of low-pass 8th Chebyshev Type II filter with 40dB of attenuation and an 8Hz cutoff frequency; (III) Down-sampling to 34.133Hz to obtain 1,024 samples per window; (IV) Normalization of the signal based on clipping at 3 standard deviations followed by a z-score standardization.
After preprocessing, the PPG signal of each subject was split into the corresponding 30-second windows to match the sleep staging labels.
This windowing is depicted in Figure~\ref{fig:ppg_split}.
In the MESA dataset, subjects have varying PPG recording lengths, resulting in different numbers of windows assigned to each subject.

\subsection{Super Windows} \label{sub_sec:super_windows}
\begin{figure}[!htb]

    \begin{subfigure}{\textwidth}
      \centering
      \includegraphics[width=.6\textwidth]{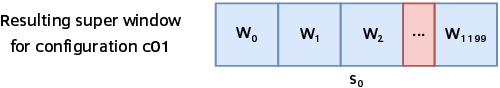}
      \caption{Super window built with configuration c01.}
      \label{fig:super_window_c01}
    \end{subfigure}
    \vspace{5mm}

    \begin{subfigure}{\textwidth}
      \centering
      \includegraphics[width=\textwidth]{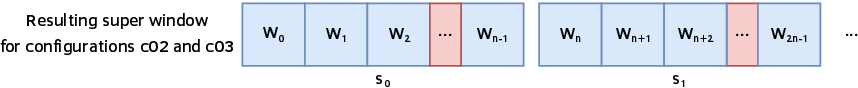}
      \caption{Super-windows built with configurations c02 and c03. n is the number of windows per super-window}
      \label{fig:super_window_c02_c03}
    \end{subfigure}
    \vspace{5mm}

    \begin{subfigure}{\textwidth}
      \centering
      \includegraphics[width=.6\textwidth]{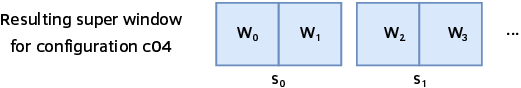}
      \caption{Super-window built with configuration c04.}
      \label{fig:super_window_c04}
    \end{subfigure}
    \vspace{5mm}

    \begin{subfigure}{\textwidth}
      \centering
      \includegraphics[width=.9\textwidth]{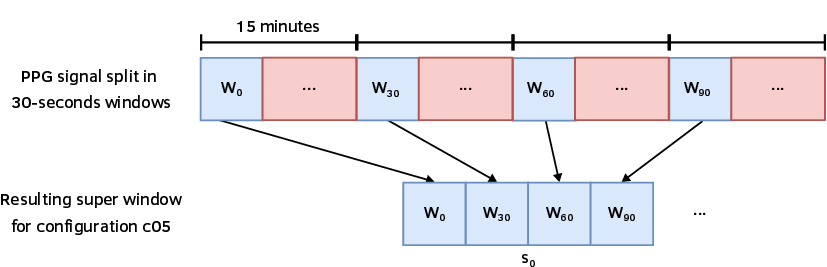}
      \caption{Super-window built with configuration c05.}
      \label{fig:super_window_c05}
    \end{subfigure}

  \caption{Block representations super windows. Each window W represents 30 seconds of PPG signal.}
  \label{fig:ppg_sw}
\end{figure}

Classification models based on raw PPG signals receive a time segment of PPG as input and return the sleep stage for each window within the time segment.
This segment can range from minutes to hours of PPG signal.
An increased number of windows allows the classification model to utilize more temporal context information.
In this work, we generalize the concept of the time segment into an arrangement of PPG signal windows.
Additionally, we propose five configurations (c01 to c05) to build these super-windows for comparison (Figure~\ref{fig:ppg_sw}).

The first configuration, c01, depicted in Figure~\ref{fig:super_window_c01}, represents the input used in \cite{kotzen_sleepppg-net_2023,nam_insightsleepnet_2024}.
It defines a continuous 10-hour (1,200 30-second windows) PPG signal as a single super-window per subject.
For subjects with PPG signals exceeding 1,024 windows, only the first 1,024 windows are used.
Conversely, for subjects with PPG signals containing fewer than 1,024 windows, the signal is padded with a zero vector to reach the 10-hour duration.

To evaluate the impact of less temporal context information in the model, the configurations c02 and c03 were defined.
Similar to c01, these configurations consist of concatenated continuous windows, as depicted in Figure~\ref{fig:super_window_c02_c03}.
However, unlike c01, configurations c02 and c03 allow for more super-windows per subject, thus maximizing the use of the PPG signal for each individual.
The primary difference between c02 and c03 lies in the number of windows composing the super-window.
In Figure~\ref{fig:super_window_c02_c03}, configuration c02 has $n$ equal to 120 windows, corresponding to 1 hour of PPG signal, whereas configuration c03 has $n$ equal to 60 windows, resulting in 30 minutes of PPG signal.

Configuration c04, represented in Figure~\ref{fig:super_window_c04}, was defined to evaluate the extreme case of using only one minute of PPG signal as input to the model.
In this case, the super-window has only 2 windows.

Wearables have limited battery life constraints, making them capable of recording short intervals.
To emulate a wearable recording PPG signal for 30 seconds every 15 minutes, we defined configuration c05 depicted in Figure~\ref{fig:super_window_c05}.
This proposal uses the temporal context information of 1 hour of PPG signal by picking one window every 15 minutes, creating a super-window with 4 windows.
Additionally, to maximize the use of the MESA dataset, we built overlapping super-windows, generating one for each adjacent window.

\subsection{SleepPPG-Net adaptation} \label{sub_sec:sleepppgnet_adaptation}
SleepPPG-Net is a convolutional neural network (CNN) model for sleep staging classification using raw PPG signals \cite{kotzen_sleepppg-net_2023}.
This model receives a 10-hour 1D PPG signal as input and outputs the sleep stage of each of the 1,200 windows.
The model is composed of three fundamental blocks: (I) Window feature extraction block --, this block is composed of various 1D residual convolutional layers that extract features from the PPG signal.
The resulting tensor is then reshaped to rearrange features per window, followed by a dense layer; (II) The temporal context information addition block -- this block includes various 1D residual dilated convolutional layers that mix features from multiple windows to incorporate temporal context information; (III) Classifier block -- this block consists of a 1D convolutional layer with 4 filters, one for each sleep stage, followed by a softmax activation function.

\begin{figure}[htb]
    \begin{subfigure}{.45\textwidth}
        \centering
        \includegraphics[width=.68\textwidth]{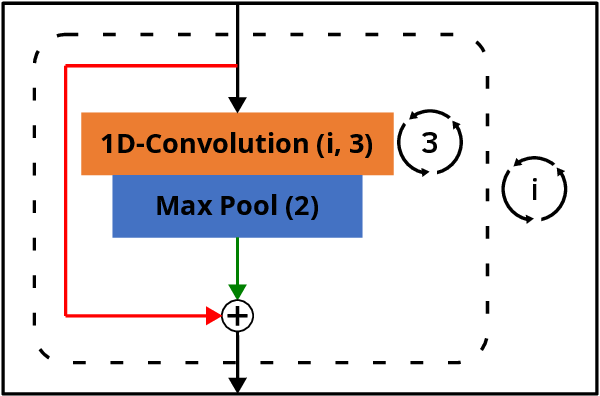}
        \caption{Original feature extraction.}
        \label{fig:original_fe}
    \end{subfigure}%
    \begin{subfigure}{.45\textwidth}
        \centering
        \includegraphics[width=\textwidth]{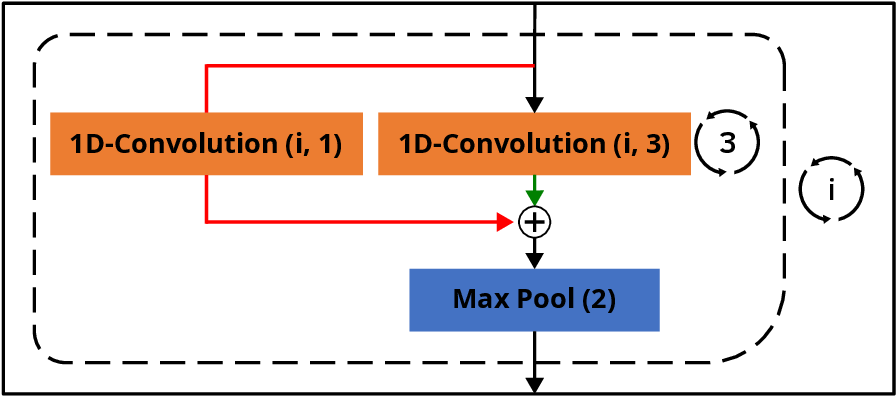}
        \caption{Adapted feature extraction.}
        \label{fig:adapted_fe}
    \end{subfigure}%
    \caption{Feature extraction in the original and adapted model. The red arrow indicates skip connection and the green arrow indicates convolution output.}
    \label{fig:adaptation}
\end{figure}

\begin{figure}[htb]
  \centering
  \includegraphics[width=0.7\textwidth]{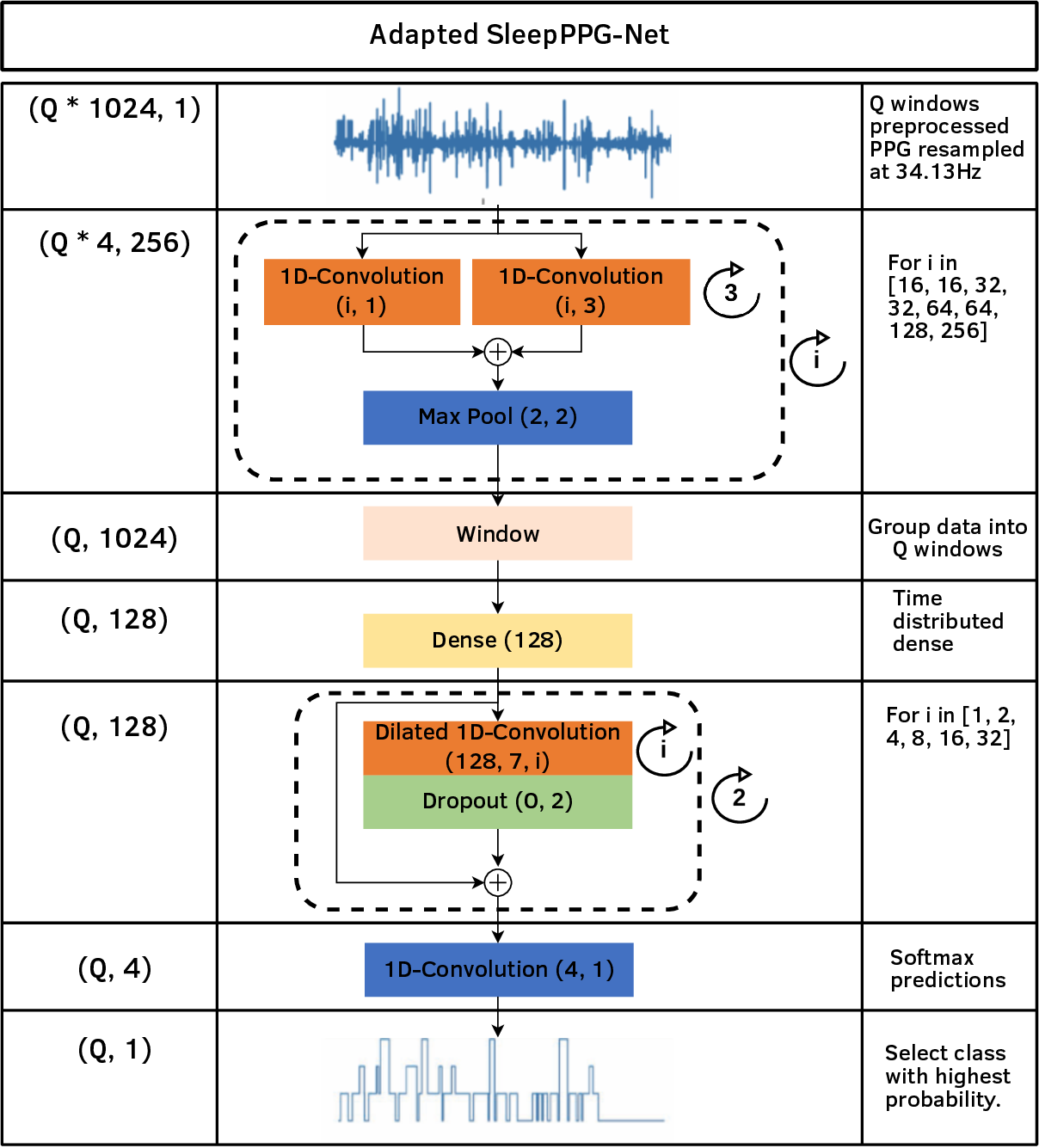}
  \caption{Based on \cite{kotzen_sleepppg-net_2023}. Diagram of the adapted SleepPPG-Net model. $Q$ is the number of windows in the input.}
  \label{spn_diagram}
\end{figure}

The feature extraction block of the original SleepPPG-Net is shown in Figure~\ref{fig:original_fe}.
This block applies a residual connection, which consists of the summing of the input (red arrow) with the output of the three convolutions followed by max pooling (green arrow).
In the SleepPPG-Net, the input tensor of the feature extraction block is a rank-2 tensor of dimension $(N, M)$.
After the three convolutions and the max pooling, the dimensions change to $(N/2, i)$, making it impossible to sum with the input $(N, M)$.
To resolve this, as shown in Figure~\ref{fig:adapted_fe}, we inserted a 1D-convolution with a kernel size of 1 in the skip connection (red arrow) and moved the max-pooling out of the residual connection.
This modification enables the summing of two tensors of dimension $(N, i)$.
The resulting adapted SleepPPG-Net model is depicted in Figure~\ref{spn_diagram}.

\subsection{Training and Evaluation} \label{sub_sec:training_and_evaluation}
To evaluate the proposal against the SOTA benchmark using the MESA Sleep dataset, the labels were converted.
The NREM1 and NREM2 were merged into a single class, resulting in a four-class classification: Wake (Wakefulness), Light (NREM1+NREM2), Deep (NREM3), and REM (REM).

The evaluation of the models was conducted using a 5-fold cross-validation.
The metrics used to evaluate the models were accuracy (Acc), Cohen’s kappa (Kappa), weighted F1 (F1-W), and macro F1 (F1-M).
Accuracy is the ratio of agreement between the predicted and true labels.
Kappa measures agreement between the predicted and true labels, accounting for agreement by chance.
The F1 score is the harmonic mean of the precision and recall.
The F1-W is the weighted average of the F1 score per class, giving more importance to over-represented classes, as used in \cite{nam_insightsleepnet_2024}.
F1-M is the average of the F1 score per class, equally representing all classes.
Some windows in the MESA dataset lack labels, so these labels were ignored in metrics and loss calculations, as were the concatenated zero windows in the super-windows.

The implementation used Python 3.7 and TensorFlow 2.3.
As in \cite{kotzen_sleepppg-net_2023}, the models were trained using the Adam optimizer with a learning rate of 0.00025, a seed set to 42, categorical cross-entropy loss, and 30 epochs.
The batch sizes were 8, 32, 64, 1,024, and 1,024 for the configurations c01, c02, c03, c04, and c05, respectively.

\section{Results}\label{sec:results}

Table~\ref{tab:results} presents the comparison of the adapted SleepPPG-Net model with different configurations against other SOTA models.
Experiments 1 and 2, using results from original papers, compare methods that utilize 1,200 windows (10 hours) of PPG signal: SleepPPG-Net \cite{kotzen_sleepppg-net_2023} and InsightSleepNet with 0.8 energy threshold \cite{nam_insightsleepnet_2024}, achieved accuracies of 0.83 and 0.86, respectively.
Experiments 3 to 7, represent evaluations conducted in this work with the adapted SleepPPG-Net model under various configurations.
Additionally, experiment 8 includes a method that uses only 6 minutes of PPG signal, SLAMSS \cite{song_ai-driven_2023} which achieved an accuracy of 0.72 in the original paper.
Furthermore, we re-implemented the XG-Boost-based classifier from \cite{almeida_machine-learning_2024}.
Originally, the model used the ACT as input (a variable derived from the actigraphy), Heart Rate (HR), and Heart Rate Variation (HRV), both derived from the PPG signal.
In this work, the model was trained using only HR and HRV.
While the original model aimed at sleep detection (a 2-class classification: Wake/Sleep), it was trained here to classify the 4 sleep stages.

\begin{figure}[!htb]

\begin{subfigure}{.5\textwidth}
  \centering
  \includegraphics[width=1.0\textwidth]{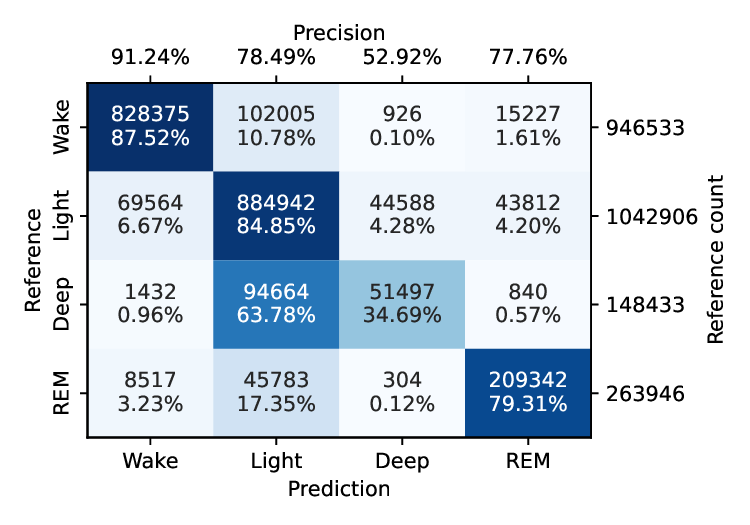}
  \caption{Configuration c01}
  \label{fig:conf_matrix_c01}
\end{subfigure}%
\begin{subfigure}{.5\textwidth}
  \centering
  \includegraphics[width=1.0\textwidth]{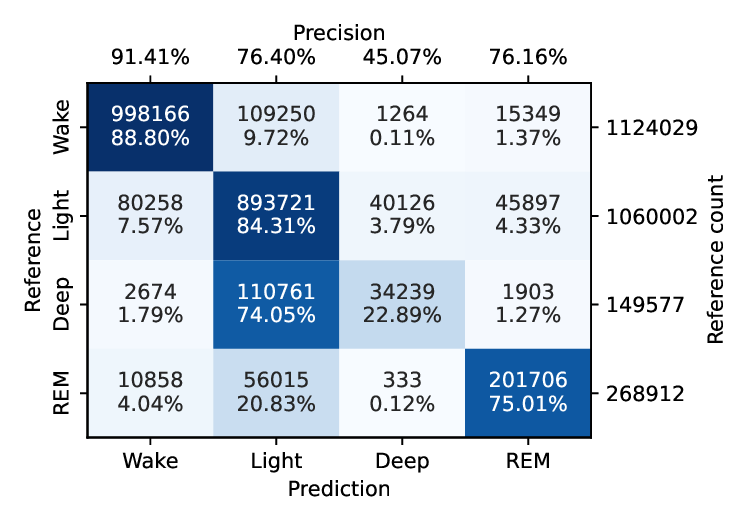}
  \caption{Configuration c02}
  \label{fig:conf_matrix_c02}
\end{subfigure}

\begin{subfigure}{.5\textwidth}
  \centering
  \includegraphics[width=1.0\textwidth]{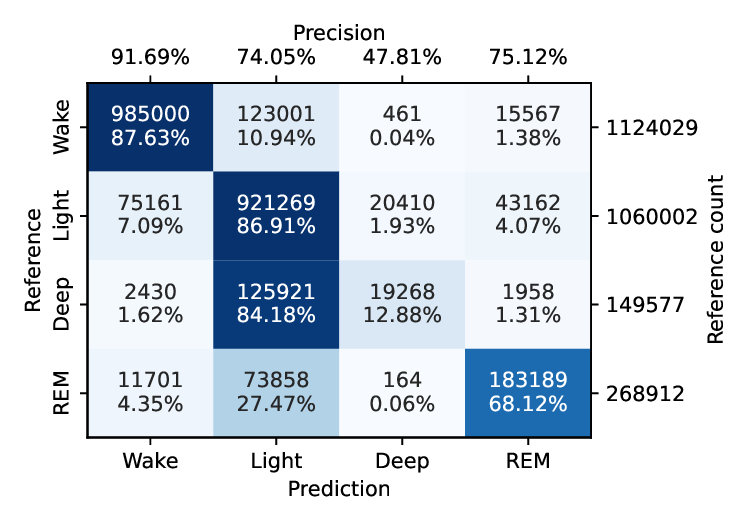}
  \caption{Configuration c03}
  \label{fig:conf_matrix_c03}
\end{subfigure}%
\begin{subfigure}{.5\textwidth}
  \centering
  \includegraphics[width=1.0\textwidth]{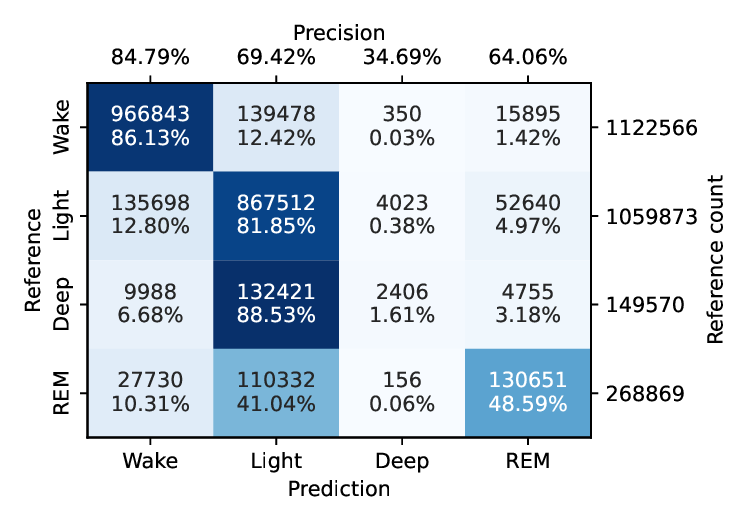}
  \caption{Configuration c04}
  \label{fig:conf_matrix_c04}
\end{subfigure}

\begin{subfigure}{.5\textwidth}
  \centering
  \includegraphics[width=1.0\textwidth]{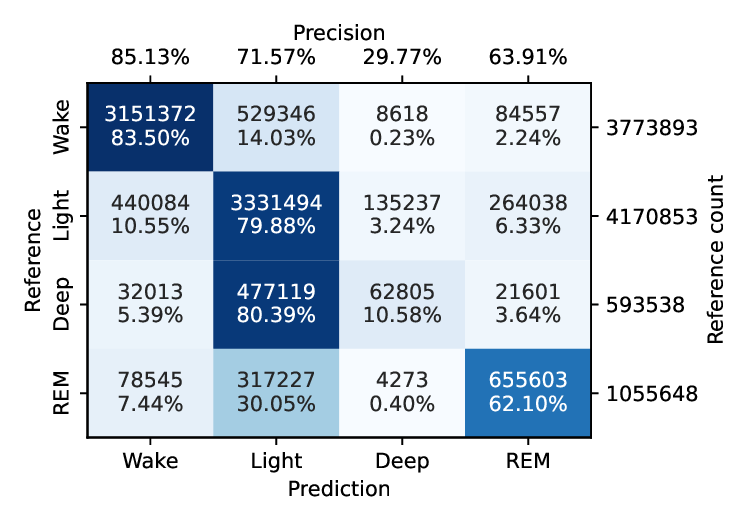}
  \caption{Configuration c05}
  \label{fig:conf_matrix_c05}
\end{subfigure}

\centering
\caption{Confusion matrix of the adapted SleepPPG-Net with different configurations.}
\label{fig:conf_matrix}
\end{figure}

To provide more details about the performance of the adapted SleepPPG-Net model with the given configurations, we present the set of confusion matrices in Figure~\ref{fig:conf_matrix}.

\begin{table}[!htb]
\centering
\resizebox{\textwidth}{!}{%
\begin{tabular}{|c|c|c|c|c|c|c|}
\hline
\textbf{E} & \textbf{Model} & \textbf{Input} & \textbf{Acc} & \textbf{Kappa} & \textbf{\begin{tabular}[c]{@{}c@{}}F1-W\end{tabular}} & \textbf{\begin{tabular}[c]{@{}c@{}}F1-M\end{tabular}} \\ \hline

1 & \cite{kotzen_sleepppg-net_2023} & 10h & $0.83$ & $0.74$ & - & - \\ \hline

2 & \begin{tabular}[c]{@{}c@{}} \cite{nam_insightsleepnet_2024}\end{tabular}  & 10h & $0.86$           & $0.78$          & $0.86$          & - \\ \hline

3 & \begin{tabular}[c]{@{}c@{}}c01-Adapted \\ SleepPPG-Net\end{tabular}       & 10h & $0.82~\pm0.004$  & $0.72~\pm0.006$ & $0.82~\pm0.005$ & $0.73\pm0.1828$ \\ \hline

4 & \begin{tabular}[c]{@{}c@{}}c02-Adapted \\ SleepPPG-Net\end{tabular} & 1h & $0.82~\pm0.003$ & $0.71~\pm0.004$ & $0.81~\pm0.004$ & $0.69~\pm0.2294$ \\ \hline

5 & \begin{tabular}[c]{@{}c@{}}c03-Adapted \\ SleepPPG-Net\end{tabular} & 30m & $0.81~\pm0.005$ & $0.79~\pm0.009$ & $0.80~\pm0.007$ & $0.65~\pm0.2678$ \\ \hline

6 & \begin{tabular}[c]{@{}c@{}}c04-Adapted \\ SleepPPG-Net\end{tabular} & 1m & $0.76~\pm0.006$ & $0.60~\pm0.010$ & $0.73~\pm0.006$ & $0.55~\pm0.3173$ \\ \hline

7 & \begin{tabular}[c]{@{}c@{}}c05-Adapted \\ SleepPPG-Net\end{tabular} & 4 $\times$ 30s & $0.75~\pm0.005$ & $0.60~\pm0.009$ & $0.74~\pm0.005$ & $0.60~\pm0.2650$ \\ \hline

8 & \cite{song_ai-driven_2023} & 6m & $0.72$ & - & $0.73$ & - \\ \hline
9 & \begin{tabular}[c]{@{}c@{}}Adapted \\ \cite{almeida_machine-learning_2024}\end{tabular} & 30s & $0.60$ & $0.30$ & $0.54$ & - \\ \hline
\end{tabular}}
\caption{Performance comparison of the adapted SleepPPG-Net model with the given configurations against other SOTA models. Where E is experiment, Acc is accuracy, Kappa is Cohen's Kappa, F1-W is weighted F1 score, and F1-M is macro F1 score.}
\label{tab:results}
\end{table}

\section{Discussion}\label{sec:discussion}

As shown in Table~\ref{tab:results}, in experiment 3, the performance of the Adapted SleepPPG-Net model with configuration c01, the same configuration used in the original SleepPPG-Net article, is only one point lower in accuracy and two points lower in Cohen’s Kappa than the original article, shown in experiment 1.
This difference can be attributed to the pretraining done in the original model with the SHHS dataset \cite{zhang_national_2018,quan_sleep_1997}, which was not performed here.
However, the performance remains comparable to the best models in the literature, shown in experiments 1 and 2.
The confusion matrix in Figure~\ref{fig:conf_matrix_c01} demonstrates that the model’s performance is similar to that reported by \cite{kotzen_sleepppg-net_2023}.

In experiments 4 and 5, the Adapted SleepPPG-Net was tested with 1 hour (c02) and 30 minutes (c03) of PPG signal length, respectively.
The accuracy, Cohen’s Kappa, and F1-weighted do not decrease significantly, but the F1-macro decreases more substantially.
This suggests that the model finds it challenging to correctly classify the class with the fewest, deep-sleep (N3).
This is confirmed by the sensitivity of the deep class in Figure~\ref{fig:conf_matrix_c02} for c02 and Figure~\ref{fig:conf_matrix_c03} for c03.
While the sensitivity for wake and light classes remains consistent, the sensitivity of the REM class decreases, and the sensitivity of the deep class decreases even further.

In experiment 6, configuration c04 uses only 1 minute of PPG Signal resulting in significantly decreased accuracy, Cohen’s Kappa, and F1-weighted, with a considerable drop in F1-macro.
As shown in the confusion matrix in Figure~\ref{fig:conf_matrix_c04}, the model fails to classify the deep and REM classes, often misclassifying the deep class as light.

In experiment 7, configuration c05 concatenates 4 windows of PPG signal measured every 15 minutes, creating a downsampled version of 1 hour of PPG signal.
Compared to c04, c05 achieves one point lower in accuracy but 5 points higher in F1-Macro.
This suggests that underrepresented classes, such as deep sleep, are better classified with this configuration due to the increased temporal context information.
The confusion matrix in Figure~\ref{fig:conf_matrix_c05} confirms this, showing a significant increase in the sensitivity of the deep sleep class, and a slight increase in the sensitivity of the REM class.
Additionally, the adapted SleepPPG-Net with configuration c05, outperforms the model presented in \cite{song_ai-driven_2023}, from experiment 8, and the adapted version of \cite{almeida_machine-learning_2024}, from experiment 9.
This demonstrates that the proposed strategy effectively leverages temporal context information similar to SleepPPG-Net and InsightSleepNet while being compatible with the periodic recording of short PPG signal segments, a more feasible approach given the battery limitations of wearable devices.

\section{Conclusion}\label{sec:conclusion}
PPG-based wearables have emerged as a more affordable alternative to PSG for obtaining sleep stages.
However, current high-performance sleep staging models using PPG signals require long segments of continuous PPG recording, which is challenging for wearable devices, such as smartwatches, due to battery life constraints.
On the other hand, models using shorter PPG segments are more suitable for real applications but generally have lower performance.
In this work, we evaluated an adapted version of the SleepPPG-Net model with various PPG segment sizes.
We found that decreasing the length of the PPG signal input reduced overall performance, particularly for the deep and REM sleep stages.
We also found that using 1 hour of PPG signal resulted in performance comparable to using 10 hours, thereby reducing the segment size requirement for high-performance sleep staging.
We also proposed a novel arrangement of PPG signal that leverages the temporal context information of long segments of continuous recording while using shorter segments.
This approach, when applied to the adapted SleepPPG-Net, outperformed SOTA models that are also able to run with shorter segments of PPG signal.

Future work includes applying the strategy of reducing temporal context to the InsightSleepNet model.
Additionally, pre-training the model using the SHHS dataset, as demonstrated in \cite{kotzen_sleepppg-net_2023}, is expected to enhance performance.
Finally, we plan to evaluate the performance of the proposed configurations across different datasets to assess the generalizability of the results.

\section*{Acknowledgements}\label{sec:acknowledgements}
This work was supported by Foxconn Brazil and Zerbini Foundation as part of the research project ``Remote Patient Monitoring System''.

\bibliographystyle{sbc}
\bibliography{
  references/sleep-staging,
  references/custom
}

\begin{thebibliography}{}

\bibitem[Almeida et~al. 2024]{almeida_machine-learning_2024}
Almeida, D.~A., Dias, F.~M., Toledo, M. A.~F., Cardenas, D. A.~C., Oliveira, F. A.~C., Ribeiro, E., Krieger, J.~E., and Gutierrez, M.~A. (2024).
\newblock A machine-learning sleep-wake classification model using a reduced number of features derived from photoplethysmography and activity signals.
\newblock In {\em Simpósio {Brasileiro} de {Computação} {Aplicada} à {Saúde} ({SBCAS})}, pages 61--69. SBC.

\bibitem[Chen et~al. 2015]{chen_racialethnic_2015}
Chen, X., Wang, R., Zee, P., Lutsey, P.~L., Javaheri, S., Alcántara, C., Jackson, C.~L., Williams, M.~A., and Redline, S. (2015).
\newblock Racial/{Ethnic} {Differences} in {Sleep} {Disturbances}: {The} {Multi}-{Ethnic} {Study} of {Atherosclerosis} ({MESA}).
\newblock {\em Sleep}, 38(6):877--888.

\bibitem[Colten et~al. 2006]{colten_extent_2006}
Colten, H.~R., Altevogt, B.~M., and Research, I. o. M. U. C. o. S. M.~a. (2006).
\newblock Extent and {Health} {Consequences} of {Chronic} {Sleep} {Loss} and {Sleep} {Disorders}.
\newblock In {\em Sleep {Disorders} and {Sleep} {Deprivation}: {An} {Unmet} {Public} {Health} {Problem}}. National Academies Press (US).

\bibitem[Djanian et~al. 2022]{djanian_sleep_2022}
Djanian, S., Bruun, A., and Nielsen, T.~D. (2022).
\newblock Sleep classification using {Consumer} {Sleep} {Technologies} and {AI}: {A} review of the current landscape.
\newblock {\em Sleep Medicine}, 100:390--403.

\bibitem[Escourrou et~al. 2000]{escourrou_needs_2000}
Escourrou, P., Luriau, S., Rehel, M., Nédelcoux, H., and Lanoë, J.~L. (2000).
\newblock Needs and costs of sleep monitoring.
\newblock {\em Studies in Health Technology and Informatics}, 78:69--85.

\bibitem[Iber et~al. 2007]{iber2007aasm}
Iber, C., Ancoli-Israel, S., and Chesson~Jr, A.~L. (2007).
\newblock The aasm manual for the scoring of sleep and associated events.
\newblock {\em Westchester, IL: American Academy of Sleep Medicine}.

\bibitem[Kotzen et~al. 2023]{kotzen_sleepppg-net_2023}
Kotzen, K., Charlton, P.~H., Salabi, S., Amar, L., Landesberg, A., and Behar, J.~A. (2023).
\newblock {SleepPPG}-{Net}: {A} {Deep} {Learning} {Algorithm} for {Robust} {Sleep} {Staging} {From} {Continuous} {Photoplethysmography}.
\newblock {\em IEEE JOURNAL OF BIOMEDICAL AND HEALTH INFORMATICS}, 27(2).

\bibitem[Nam et~al. 2024]{nam_insightsleepnet_2024}
Nam, B., Bark, B., Lee, J., and Kim, I.~Y. (2024).
\newblock {InsightSleepNet}: the interpretable and uncertainty-aware deep learning network for sleep staging using continuous {Photoplethysmography}.
\newblock {\em BMC Medical Informatics and Decision Making}, 24(1):50.

\bibitem[Olsen et~al. 2023]{olsen_flexible_2023}
Olsen, M., Zeitzer, J.~M., Richardson, R.~N., Davidenko, P., Jennum, P.~J., Sorensen, H. B.~D., and Mignot, E. (2023).
\newblock A {Flexible} {Deep} {Learning} {Architecture} for {Temporal} {Sleep} {Stage} {Classification} {Using} {Accelerometry} and {Photoplethysmography}.
\newblock {\em IEEE Transactions on Biomedical Engineering}, 70(1):228--237.

\bibitem[Quan et~al. 1997]{quan_sleep_1997}
Quan, S.~F., Howard, B.~V., Iber, C., Kiley, J.~P., Nieto, F.~J., O'Connor, G.~T., Rapoport, D.~M., Redline, S., Robbins, J., Samet, J.~M., and Wahl, P.~W. (1997).
\newblock The {Sleep} {Heart} {Health} {Study}: design, rationale, and methods.
\newblock {\em Sleep}, 20(12):1077--1085.

\bibitem[Rundo and Downey 2019]{rundo_chapter_2019}
Rundo, J.~V. and Downey, R. (2019).
\newblock Chapter 25 - {Polysomnography}.
\newblock In Levin, K.~H. and Chauvel, P., editors, {\em Handbook of {Clinical} {Neurology}}, volume 160 of {\em Clinical {Neurophysiology}: {Basis} and {Technical} {Aspects}}, pages 381--392. Elsevier.

\bibitem[Song et~al. 2023]{song_ai-driven_2023}
Song, T.-A., Chowdhury, S.~R., Malekzadeh, M., Harrison, S., Hoge, T.~B., Redline, S., Stone, K.~L., Saxena, R., Purcell, S.~M., and Dutta, J. (2023).
\newblock {AI}-{Driven} sleep staging from actigraphy and heart rate.
\newblock {\em PLOS ONE}, 18(5):e0285703.

\bibitem[Zhang et~al. 2018]{zhang_national_2018}
Zhang, G.-Q., Cui, L., Mueller, R., Tao, S., Kim, M., Rueschman, M., Mariani, S., Mobley, D., and Redline, S. (2018).
\newblock The {National} {Sleep} {Research} {Resource}: towards a sleep data commons.
\newblock {\em Journal of the American Medical Informatics Association: JAMIA}, 25(10):1351--1358.

\end{thebibliography}

\end{document}